\documentclass[runningheads]{llncs}
\usepackage{graphicx}
\usepackage{color,soul}

\usepackage{hyperref} %Added by GRG

\usepackage{subcaption}

\usepackage{array} % to control column spacing in matrix of figures.
\usepackage{graphicx,multirow}

\usepackage{bbm} % for Indicator function

\makeatletter
\newcommand{\printfnsymbol}[1]{%
  \textsuperscript{\@fnsymbol{#1}}%
}
\begin{document}
\title{The Role of Pleura and Adipose in Lung Ultrasound AI}
% \title{The Pleural Line and Soft Tissue Space in Lung Ultrasound AI}
% \title{The Pleural Line and Soft Tissue Space in Lung Ultrasound AI \thanks{Supported by organization x.}}

% The Role of Pleura in Ultrasound AI for COVID-19 and other Lung Diseases 


\author{
% First Author\inst{1}\orcidID{0000-1111-2222-3333} \and
% Gautam Rajendrakumar Gare\inst{1}\orcidID{0000-0002-1689-9626}\thanks{equal contribution} \and
% % Wanwen Chen\inst{1}\orcidID{0000-0001-5067-5223}\footnotemark[1] \and
% Wanwen Chen\inst{1}\orcidID{0000-0001-5067-5223}\printfnsymbol{1} \and
% Alex Ling Yu Hung\inst{1}\orcidID{0000-0002-2664-7430} \and
% Edward Chen\inst{1}\orcidID{} \and
% Hai V. Tran\inst{2}\orcidID{} \and 
% Tom Fox\inst{2}\orcidID{} \and 
% Pete Lowery\inst{2}\orcidID{} \and 
% Kevin Zamora\inst{2}\orcidID{} \and 
% Bennett P deBoisblanc\inst{2}\orcidID{} \and 
% Ricardo Luis Rodriguez\inst{3}\orcidID{} \and
% John Michael Galeotti \inst{1}\orcidID{0000-0003-4247-4311}
Gautam Rajendrakumar Gare\inst{1}\thanks{equal contribution; corresponding author gautam.r.gare@gmail.com} \and
% Wanwen Chen\inst{1}\orcidID{0000-0001-5067-5223}\footnotemark[1] \and
Wanwen Chen\inst{1}\printfnsymbol{1} \and
Alex Ling Yu Hung\inst{1} \and
Edward Chen\inst{1} \and
Hai V. Tran\inst{2} \and 
Tom Fox\inst{2} \and 
Pete Lowery\inst{2} \and 
Kevin Zamora\inst{2} \and 
Bennett P deBoisblanc\inst{2} \and 
Ricardo Luis Rodriguez\inst{3} \and
John Michael Galeotti \inst{1}
}

% %
\authorrunning{G. Gare et al.}
\institute{Robotics Institute and Department of ECE, Carnegie Mellon University, USA \and
Dept. of Pulmonary and Critical Care Medicine, Louisiana State University, USA \and
Cosmeticsurg.net, LLC, Baltimore, USA\\
% \email{gautam.r.gare@gmail.com}
}

\maketitle              % typeset the header of the contribution
\begin{abstract}
% \textbf{The abstract should briefly summarize the contents of the paper in 15--250 words.}

In this paper, we study the significance of the pleura and adipose tissue in lung ultrasound AI analysis. We highlight their more prominent appearance when using high-frequency linear (HFL) instead of curvilinear ultrasound probes, showing HFL reveals better pleura detail. We compare the diagnostic utility of the pleura and adipose tissue using an HFL ultrasound probe. Masking the adipose tissue during training and inference (while retaining the pleural line and Merlin's space artifacts such as A-lines and B-lines) improved the AI model's diagnostic accuracy. 

% we establish that the pleura viewed as the pleural line, A-line, and B-line artifacts in ultrasound images hold significant diagnostic characteristics for AI analysis of lung ultrasound.     

% Our experiments, establish that the pleura viewed as the pleural line, A-line, and B-line artifacts in ultrasound images hold significant diagnostic characteristics for AI analysis of lung ultrasound.

% In this paper, we reiterate the significance of the pleura in Pulmonary Ultrasound AI.

% \keywords{Pulmonary Ultrasound  \and Pleura \and Adipose \and Deep Learning \and Machine Learning \and Linear Probe \and Lung \and Pleural Line.}

% \keywords{Lung Ultrasound \and Pleura \and Adipose \and Deep Learning \and Linear Probe \and Pulmonary Diseases }


\end{abstract}
%
%
%

% \textbf{Page limit -7 : + 2 ref}

\section{Introduction}
% % Covid and bed side monitoring benefit using Ultrasound probe
% % Prior to COVID, Neonatal Lung ultrasound AI - Check if linear or convex probe used?
% % Given this large research done \cite{} in US AI

Point-of-care ultrasound (POCUS) is a non-invasive, real-time, bedside patient monitoring tool that is ideal for working with infectious diseases such as COVID-19. POCUS does not require the transport of critically ill contagious patients to a radiology suite, making POCUS an easy choice for obtaining serial imaging to closely monitor disease progression. This has led to a significant interest in developing AI approaches for the interpretation of lung ultrasound (LUS) imaging \cite{Born2020POCOVID-net:POCUS,Roy2020DeepUltrasound,Zhang2020LungCOVID-19b}.

A healthy lung is filled with air, which results in poor visibility of the internal anatomy in ultrasound images. Typical clinical practice for pulmonary ultrasound does not try to image the internal tissue of the lung but rather focuses on artifacts (e.g., ``B Lines'') that are physically generated at the pleural membrane line. Traditionally, lung ultrasound in the ICU uses a low frequency (1-5 MHz) curvilinear or phased array (i.e., echocardiography) probe, which provides relatively deep and wide imaging. This approach is excellent for penetrating soft-tissues superficial to the lung and is standard practice for detecting prominent B-lines. Unsurprisingly, most AI research has followed typical clinical practice and focused on curvilinear probes and B-lines \cite{Roy2020DeepUltrasound,Xue2021ModalityInformation}. 
However, low-frequency curvilinear probes provide very poor detail of the pleural line. High-frequency-linear (HFL) ultrasound probes (typically in the 5-15 MHz range) offer higher resolution of the pleural line but they can not image below 6-10 cm. Such reduced imaging depth is not a fundamental problem for lung ultrasound, except in cases of obesity or deep lung consolidation.  The wavefronts generated from a linear array probe travel in parallel and generally intersect the visceral pleura in the perpendicular direction. The acoustic reflection is then displayed as a rectangular image. In contrast, the acoustic wavefront from a curvilinear probe travels radially and strikes the visceral pleura at various incident angles. %In theory, curvilinear probes could be advantages over linear probes by permitting imaging behind ribs and by identifying reverberations that are unique to the incident angle of the wavefront.
In Fig.~\ref{fig:curvilinear_vs_linear} we observe that on the curvilinear probe more depth is visible whereas on the linear probe the pleural line details are better seen. HFL probes are especially suitable for the easily-accessible L1 and R1 viewpoints, where their more detailed picture of the shallow lung's visceral pleural membrane can have diagnostic value.

Precise pleural line imaging may be more useful than deeper imaging penetration, even for pulmonary diseases that do not primarily manifest in the pleura. For example, Carrer et al.'s automated method \cite{Carrer2020AutomaticDatab} to extract the pleural line and assess lung severity achieved better results from a linear probe compared to those from a curvilinear probe.  Neonatal lung ultrasound analysis, which has also been extensively researched \cite{Liang2018UltrasoundDisease}, typically calls for an HFL probe for its improved lung surface image quality.  Although recent large datasets on COVID-19 \cite{Born2021AcceleratingAnalysis,Roy2020DeepUltrasound,Xue2021ModalityInformation} predominantly contain images from lower frequency curvilinear and phased array probes, they also include a small portion of linear probe images based on their observation that a linear probe is better for pleura visualization \cite{Born2021AcceleratingAnalysis,Soldati2020IsPandemic}.

% \textbf{[Dr. Ben/ Dr. Ricardo] Significance of Pleural line vs the artifacts:} see  intro above

% Covid specific papers highlighting linear probe vs convex
Since the HFL probe provides a better view of the pleural line, it can be a better option for Covid-19 lung ultrasound diagnosis. 
The SARS-CoV-2 virus that causes Covid-19 binds to ACE-2 receptors of epithelial cells lining the bronchi and alveoli, and endothelial cells lining the pulmonary capillaries. The lung injury from COVID-19 involves interalveolar septae that perpendicularly abut the visceral pleura. Therefore the pleural line should be a focus of the investigation. B-lines, which radiate deeply below the pleural line, have been extensively reported on. Although they are visualized below the pleural line, they are merely reverberation artifacts that emanate from within the pleural line. 
Thickening and disruption of the pleural line are subtle signs of underlying lung pathology that are poorly visualized using a curvilinear probe. In some cases these pleural line abnormalities are seen in the absence of B-lines and, therefore, such cases might be misclassified if a curvilinear probe were used. We propose that there are important and clinically relevant anatomic details visible in HFL images that are lacking in curvilinear images. Focusing on the pleura line itself, where pathology manifests the earliest, may yield clinically relevant information more directly than limiting interpretation to artifact appearance.

%  We propose that there is important and clinically relevant anatomic detail visible in HFL images that is lacking in curvilinear images. Prior techniques \cite{} have hinted at this. Focusing on the pleura line itself, where pathology manifests earliest, may yield clinically relevant information more directly than limiting interpretation to artifact appearance.

% \textbf{[Dr. Ben/ Dr. Ricardo] Significance of Subcutaneous Region for pulmonary diagnosis:} 
When performing lung ultrasound with any probe, acoustic waves are first propagated through the keratinized layer of the outer skin, i.e. the epidermis. The sound then travels through fibrous tissue and capillary networks in the deeper dermis and then through adipose tissue and muscle bundles covered in the fibrous fascia. The acoustic wavefront must then traverse the 1-4 micron thick fibrous parietal pleura which lines the inside of the chest cavity before the lung's visceral pleura is reached.  The acoustic characteristics of each of these structures is affected by probe location and patient characteristics, including age, sex, anatomy, lean body mass, and fat mass.  Making use of the linear probe helps see not only the pleura but also the subcutaneous (SubQ) tissue structure in detail, which would otherwise occupy relatively few pixels with a curvilinear probe. This brings in additional challenges where a purported lung-AI network could instead rely on the SubQ to make the diagnosis rather than lung regions. AI might learn associations between these soft tissue structures and specific disease characteristics. For example, it has been well established that obesity and older age are risk factors for severe COVID. It would therefore be important that training and testing of AI approaches to the diagnosis of lung diseases consider the impact of the subcutaneous tissues on diagnostic accuracy.

We can broadly categorize the various regions that constitute the linear probe ultrasound image into the subcutaneous region, the pleura, and Merlin's space (i.e., real and artifact pixels beneath the pleural line) \cite{Lichtenstein2017NovelNow}. In the following sections of the paper, we try to determine the diagnostic prowess of these regions by generating images that emphasize these regions by masking out other regions. We study the diagnostic ability of the subcutaneous region (\emph{subq}), subcutaneous+pleura (\emph{subq+pleura}), the pleural region (\emph{pleural}), the Merlin's region (\emph{merlin}), and the pleural+Merlin's region (\emph{pleural+merlin}). In addition, we also explore masking out indirect adipose/obesity information implicitly encoded by the depth and curvature of the pleura. So, we straighten the overall bend of the pleural line and mask out the depth by shifting up the pleura to a fixed distance from the top of the image. Refer to Fig.~\ref{fig:gradcam_results} for sample masked images.

\section{Methodology}

\subsubsection{Problem Statement}
Given an ultrasound B-mode scan clip $I_g$, the task is to find a function $F \colon [ \, I_g] \, \to L$ that maps the clip $I_g$ to ultrasound severity score labels $L \in \{0, 1, 2, 3\}$ as defined by \cite{Roy2020DeepUltrasound}. Because the pleural line produces distinct artifacts (A-lines, B-lines) when scattering ultrasound based on the lung condition, the classification model should learn underlying mappings between the pleural line, artifacts, and pixel values, for making the predictions.  
% The severity scores presently being considered are: (0) score-0, (1) score-1, (2) score-2 and (3) score-3.

\subsection{SubQ Masking}
% The subcutaneous tissue (SubQ) layer may play a positive or negative role in the interpretation of lung ultrasound images. On the positive side, subcutaneous tissue involvement in the disease process may yield additional diagnostic clues. On the negative side it may generate artifacts that obscure the interpretation of subtle pleural details.

% SubQ is the abbreviation of subcutaneous, meaning ``under the skin''. 

In lung ultrasound images, the SubQ tissue has more complicated tissue structures than those in the lung region. However, these structures might degrade the performance of AI-based diagnosis. The brighter and more complicated SubQ region has a larger response to the CNN layers than the lung region, but it does not provide much information on the underlying lung diseases and might even interfere with the performance of the deep neural network. 
To understand the role of the pleura and the adipose in the COVID-19 diagnosis, we provide different masking of the lung ultrasound images. The ultrasound image is divided into SubQ, pleural line, and Merlin region.

\subsubsection{Pleural Line Segmentation}

The pleural line separates the SubQ tissue and Merlin's region in the ultrasound images and is usually the lower bright and wide horizontal line in the ultrasound images. To segment the pleural line, we first use a 5x5 Gaussian filter to blur the image and reduce the speckle noise. We then resize it to 150x150 to reduce the influence of the speckles in the segmentation. To find the candidate pixels that belong to a bright horizontal line, we first threshold the image based on the image's response to the Sobel filter along the y-axis with a 3x3 kernel size, and then threshold the image based on the intensity. We select the thresholds in Eq. \ref{eq:sobel_threshold} and Eq. \ref{eq:intensity_threshold} after tuning. Then in each column, we keep the lowest candidate point in the image, use dilation to fill the gap between the line, and then cluster the pixels into different regions based on the connectivity. We then keep the region that has the largest area, and move other regions to the same level as the largest region. This is done by adding an offset to the y coordinates of the candidate pixels in other regions. The offset is calculated by the difference between the minimal y coordinates of the region and the one of the largest region. We then fit a fourth-order polynomial curve using the candidate pixels and then extend the polynomial curve segment for more than 10 pixels along the tangent line at the two endpoints of the polynomial curve. 

\begin{equation}\label{eq:sobel_threshold}
    threshold_{sobel} = 0.2 \times mean(I_{sobel})
\end{equation}
\begin{equation}\label{eq:intensity_threshold}
    threshold_{intensity} = mean(I) + 1.3\times std(I)
\end{equation}
where $I$ is the raw image, and $I_{sobel}$ is the response of the image to Sobel filter.

With the segmented pleural line, the region above this line is the selected SubQ region, and the region below this line is the selected Merlin's region.

\subsubsection{Pleural Line Straightening}
We straighten and shift up the pleura in order to mask out the adipose/obesity information indirectly encoded into the curvature and depth of the pleura. Besides, different probe pressure would create different appearances of pleural lines in the images, so we want to eliminate the effect of this arbitrary variable as well. Therefore, we straighten the pleural lines while maintaining the local ``bumps" on the pleural lines so that local pleura information would not be lost. In practice, we crop the images at 5 pixels above the pleural lines to preserve the information on the pleural lines and underneath them. We take the upper boundaries of the segmented pleural lines and fit a cubic function to it. (We did not use a higher-order function since we would like to preserve the local information on the pleural lines.) We then shift each column of the image upwards or downwards so that we make the cubic curve into a horizontal straight line. (Refer to Fig.~\ref{fig:gradcam_results} for sample straightened image.)

\vspace{-1.0em}

\subsection{Data}

% We complied a lung ultrasound dataset sourced from the public POCOVID-Net dataset \cite{Born2020POCOVID-net:POCUS} and the custom data acquired by us from the Louisiana State University Health Science Center. 
% the publicly usable subset of the POCOVID-Net dataset \cite{Born2020POCOVID-net:POCUS, Born2021AcceleratingAnalysis}, as well as

Under IRB approval, we curated our own lung ultrasound dataset consisting primarily of linear probe videos. Our dataset consists of multiple ultrasound B-mode scans of L1 and R1 (left and right superior anterior) lung regions at depths ranging from 4cm to 6cm under different scan settings, obtained using a Sonosite X-Porte ultrasound machine. The dataset consists of ultrasound scans of 93 unique patients from the pulmonary ED during COVID-19, and some patients were re-scanned on subsequent dates, yielding 210 videos. 
% Fig.~\ref{fig:Lung-dataset} shows the data distribution into the various ultrasound severity scores and probes.

% The combined dataset consists of ultrasound scans of 93 (healthy and COVID-19) patients, totaling 210 videos (113 Healthy and 175 COVID-19). Fig.~\ref{fig:Lung-dataset} shows the data distribution into the various ultrasound severity scores and probes.

% Generating data split : data/LSU-Lung-Severity-Dataset/dataset_split_equi_class_R1.json
% 0 : Total no of Videos = 27; [Train = 19; Val = 3; Test = 5
% 1 : Total no of Videos = 84; [Train = 59; Val = 8; Test = 17
% 2 : Total no of Videos = 75; [Train = 52; Val = 8; Test = 15
% 3 : Total no of Videos = 24; [Train = 17; Val = 2; Test = 5

% The POCOVID-Net dataset which has linear and curvilinear ultrasound images collected from multiple sources with explicit CC license (excluding the \href{https://www.disi.unitn.it/iclus}{ICLUS-DB} \cite{Roy2020DeepUltrasound} data)
% Approval for acquisition and usage of our custom datasets was granted by the institutional review board of Louisiana State University Health Science Center (April 30th, 2020/No. 00000177) and Carnegie Mellon University (May 8th, 2020/ No. STUDY2020\_00000189).

% \href{https://www.butterflynetwork.com/}{butterflynetwork.com} 

% Total videos 288
% score-0 113 score-1 57 score-2 94 score-3 24

% Total frames 50363
% score-0 20319 score-1 9966 score-2 17066 score-3 3012

% Total linear 50363
% score-0 57 score-1 8 score-2 51 score-3 9

We use the same 4-level ultrasound severity scoring scheme as defined in \cite{SimpleClinicalTrials.gov} which is similarly used in \cite{Roy2020DeepUltrasound}. The score-0 indicates a normal lung with the presence of a continuous pleural line and horizontal A-line artifact. Scores 1 to 3 signify an abnormal lung, wherein score-1 indicates the presence of alterations in the pleural line with $\leq 5$ vertical B-line artifacts, score-2 has the presence of $> 5$ B-lines and score-3 signifies confounding B-lines with large consolidations (refer to \cite{Soldati2020ProposalCOVID-19} for sample images corresponding to the severity scores). All the manual labeling was performed by individuals with at least a month of training from a pulmonary ultrasound specialist. We have 27, 84, 75, and 24 videos labeled as scores 0, 1, 2, and 3 respectively. 
% Fig.~\ref{fig:gradcam_results} of 

\vspace{-1.0em}

\begin{figure}[!tbp]
  \centering
  \begin{minipage}[b]{0.57\textwidth}
  
    \begin{subfigure}{3.4cm}
    \includegraphics[width=3.4cm, height=3.5cm]{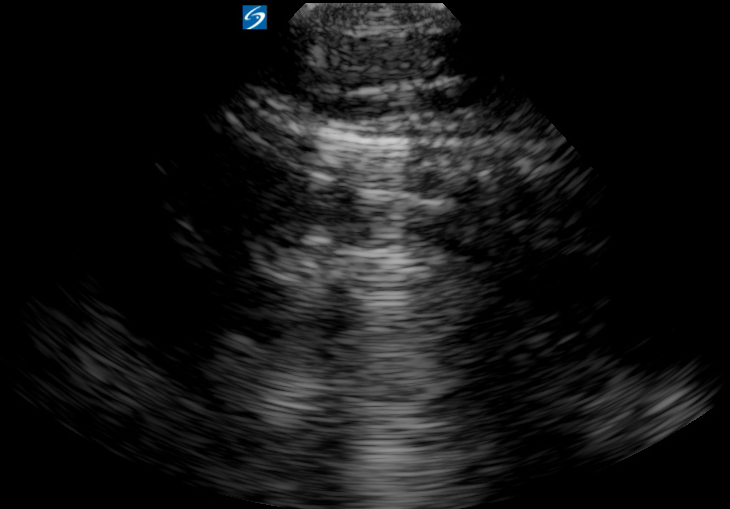} 
    % \caption{Caption1}
    % \label{fig:subim1}
    \end{subfigure}
    % \qquad
    \begin{subfigure}{3.4cm}
    \includegraphics[width=3.4cm, height=3.5cm]{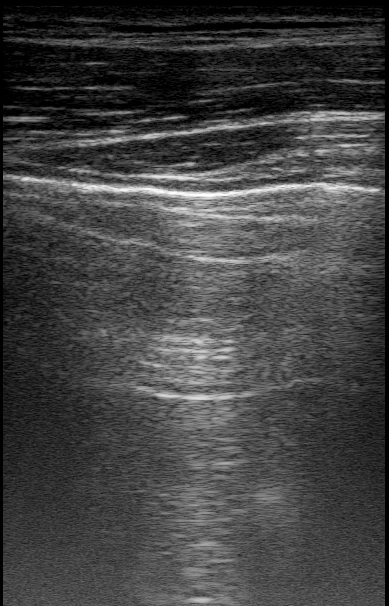}
    % \caption{Caption 2}
    % \label{fig:subim2}
    \end{subfigure}

%   \subfloat[\centering label 1]{{\includegraphics[width=3cm, height=4cm]{curvilinear_18(02.04.21)L1-P19R-313(Heroin OD with cardiac arrest)Lung_frame1.png} }}
%     \qquad
%     \subfloat[\centering label 2]{{\includegraphics[width=3cm, height=4cm]{linear_18(02.04.21)L1-HFL38R-313(Heroin OD with cardiac arrest)_frame1.png} }}
    
%     \begin{minipage}[b]{0.25\textwidth}
%     % \includegraphics[width=\textwidth]{curvilinear_18(02.04.21)L1-P19R-313(Heroin OD with cardiac arrest)Lung_frame1.png}
%     \includegraphics[width=3cm, height=4cm]{curvilinear_18(02.04.21)L1-P19R-313(Heroin OD with cardiac arrest)Lung_frame1.png}
%     % \caption{\small The.}
%   \label{fig:Lung-dataset}
%   \end{minipage}
% %   \hfill
%   \begin{minipage}[b]{0.25\textwidth}
%     \includegraphics[width=3cm, height=4cm]{linear_18(02.04.21)L1-HFL38R-313(Heroin OD with cardiac arrest)_frame1.png}
%     % \caption{\small The distribution of ultrasound video clips into various severity scores and probes.}
%   \label{fig:Lung-dataset}
%   \end{minipage}

  \caption{\small Curvilinear (left) vs Linear (right) probe at the same \textbf{L1} position during a single patient session. On the curvilinear more depth is visible whereas on the linear the pleural line details are better seen.
  }
  %   Curvilinear (left) vs Linear (right) probe on same patient and same lung region. On the curvilinear more depth is visible whereas on the linear the pleural line details is better seen.
  \label{fig:curvilinear_vs_linear}
  \end{minipage}
  \hfill
  \begin{minipage}[b]{0.4\textwidth}
    \includegraphics[width=\textwidth]{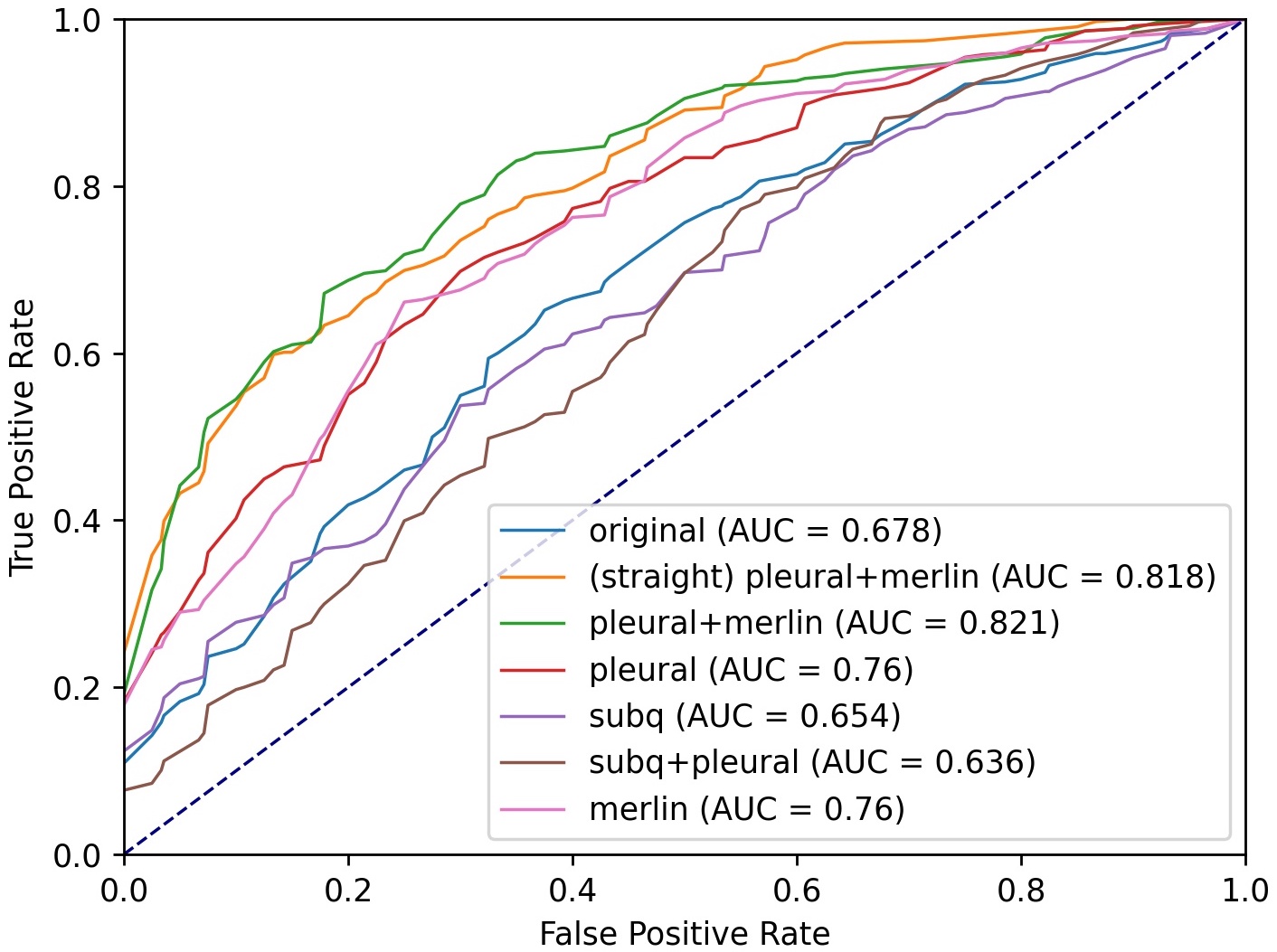}
    \caption{\small RoC plots with AUC (macro averaged) of the trained model for the video-based lung-severity scoring.}
   \label{fig:clr_roc_plot}
  \end{minipage}
\end{figure}

% \vspace{-1.0em}

\subsubsection{Data Preprocessing}

We perform dataset upsampling to address the class imbalance for the training data, wherein we upsample all the minority class labeled data to get a balanced training dataset \cite{Rahman2013AddressingDatasets}. All the images are resized to 224x224 pixels using bilinear interpolation. We augment the training data using random horizontal (left-to-right) flipping and scaling the image-pixel intensities by various scales $[0.8, 1.1]$.

\subsection{Architecture}
We carry out all our experiments on the TSM network \cite{Lin2018TSM:Understanding} with ResNet-18 (RN18) \cite{He2016DeepRecognition} backbone, commonly used for video classification and benchmarking methods. The TSM module makes use of 2D CNN's with channel mixing along the temporal direction to infuse the temporal information within the network. We use bi-directional residual shift with $1/8$ channels shifted in both directions, as recommended in \cite{Lin2018TSM:Understanding}.
% 
% The model is fed inputs clips of 18 frames wide sampled from the video by randomly selecting a start frame and then selecting 18 equally spaced frames.
% 
The model is fed input clips of 18 frames wide sampled from the video by dividing the video into 18 equal segments and then selecting an equally spaced frame from each segment beginning with a random start frame.
% dividing the video into 18 equal segments selecting a 

\subsection{Training Strategy}
% To access the ultrasound severity score of the video clips, we make use of the video labels as the noisy weak labels for the corresponding video frames. We augment the cross-entropy loss training objective for the classification task, using the contrastive learning objective in order to learn features that are robust to the frame-level label noise.

\subsubsection{Implementation}
The network is implemented with PyTorch and trained using the stochastic gradient descent algorithm \cite{Bottou2010Large-scaleDescent} with an Adam optimizer \cite{Kingma2015Adam:Optimization} set with an initial learning rate of $0.001$, to optimize over cross-entropy loss. The model is trained on an Nvidia Titan RTX GPU, with a batch size of 8 for 50 epochs. The ReduceLRonPlateau learning rate scheduler was used, which reduces the learning rate by a factor (0.5) when the performance metric (accuracy) plateaus on the validation set. For the final evaluation, we pick the best model with the highest validation set accuracy to test on the held-out test set.

% set with an initial learning rate of $0.001$, to optimize over cross-entropy loss.

\vspace{-1.0em}

\subsubsection{Metrics}
For the severity classification, we report accuracy and F1 score \cite{Born2020POCOVID-net:POCUS,Roy2020DeepUltrasound}. The receiver operating characteristic (ROC) curve is also reported along with its area under the curve (AUC) metric \cite{Kim2020ChangesStudy}, wherein a weighted average is taken where the weights correspond to the support of each class and for the multi-label we consider the one-vs-all approach. \cite{Fawcett2006AnAnalysis}

%  wherein for the calculation of the metric the weighted average is taken, where the weights correspond to the support of each class and for the multi-label we consider the one-vs-all approach.

% We evaluate semantic segmentation results using Mean Intersection over Union (mIoU) \cite{} and pixel-wise accuracy. We calculated mIoU per segmentation category and mean mIoU across all segmentation categories.

\vspace{-1.0em}

\section{Experiments}

We train the model on the various masked inputs and compare its performance to predict video-based lung-severity score labels. We randomly split the dataset into a training set and a separate held-out test set with 78\%, and 22\% split ratio respectively by randomly selecting videos while retaining the same distribution across the lung-severity scores in both the datasets and \emph{ensuring no patient overlap between the test and train set.} Using the train set, we perform \emph{5-fold cross-validation to create a training and validation fold.} The training set is upsampled to address the class imbalance \cite{Rahman2013AddressingDatasets}. We report the resulting metrics on the held-out test set in form of mean and standard deviation over the five independent cross-validation runs.
% of 46 videos (22\% of dataset)

% We train the model on the various masked inputs and compare its performance with the model trained using the original image, in order to access the robustness achieved using the various masked inputs to the predict video lung-severity score labels. 

% \emph{We conduct three independent runs, wherein each run we randomly split the dataset into train, validation, and test sets with 70\%, 10\%, and 20\% split ratio respectively, by maintaining the same split ratio for all the individual severity scored clips.} 

% \begin{table*}[!ht]
% % \begin{table*}[!h]
% \centering
% % \caption{Segmentation Pixel-wise and mIoU scores}
% \caption{Performance comparison of the video-based score prediction with other existing work.}
% % \caption{Segmentation Pixel-wise and mIoU scores obtained by evaluating on 7 independently trained models}
% \label{tab:other_methods}
% \resizebox{\textwidth}{!}{
% \begin{tabular}{|c|c|c|c|c|c|c|}
%  \hline
% Method & AUC of ROC & accuracy & precision & recall & F1-score\\
% \hline

% \cite{Roy2020DeepUltrasound}  & - & - & 0.70 & 0.60 & 0.61 \\

% \cite{Xue2021ModalityInformation} & - & 0.5660 & 0.5648 & 0.5630 & 0.5639 \\

% ours  & 0.8844 & 0.7103 & 0.7053 & 0.7103 & 0.7014 \\

% \hline
% \end{tabular}
% }
% \end{table*}

\vspace{-2.5em}

\begin{table}[!ht]
% \begin{table*}[!h]
\centering
% \caption{Segmentation Pixel-wise and mIoU scores}
\caption{Video-based 4-severity-level lung classification AUC of ROC, Accuracy, and F1 scores on a 93-patient HFL lung dataset. Highest scores shown in bold.}
% \caption{Video-based lung severity classification AUC of ROC, Accuracy, Precision, Recall, and F1 scores (weighted average) on lung dataset. Highest scores are shown in bold.}
% \caption{Segmentation Pixel-wise and mIoU scores obtained by evaluating on 7 independently trained models}
\label{tab:video_based_classification}
\resizebox{\textwidth}{!}{
\begin{tabular}{|c|c|c|c|c|}
 \hline
Method & AUC of ROC & accuracy & F1-score\\
\hline

% crop image & 0.6553 $\pm$ 0.0425 & 0.4565 $\pm$ 0.0659 &  0.4402 $\pm$ 0.0779 & 0.4565 $\pm$ 0.0659 & 0.4381 $\pm$ 0.0701Prob_based Test set : score0 precision = 0.26 +/- 0.12543258481484518 \\
% straighten pleural line &  & 0.5348 $\pm$ 0.0928 & 0.534 $\pm$ 0.029 & 0.527 $\pm$ 0.156 & 0.519 $\pm$ 0.095 \\
% under pleural line &  & 0.5261 $\pm$ 0.1178 & 0.741 $\pm$ 0.077 & 0.684 $\pm$ 0.120 & 0.707 $\pm$ 0.066 \\
% plerual line &  & 0.3783 $\pm$ 0.0525 & 0.230 $\pm$ 0.057 & \bfseries{0.360 $\pm$ 0.149} & 0.277 $\pm$ 0.085 \\ 
% subQ &  & 0.4130 $\pm$ 0.0645 & 0.230 $\pm$ 0.057 & \bfseries{0.360 $\pm$ 0.149} & 0.277 $\pm$ 0.085 \\ 
% subQ with pleural line &  & 0.3652 $\pm$ 0.0374 & 0.230 $\pm$ 0.057 & \bfseries{0.360 $\pm$ 0.149} & 0.277 $\pm$ 0.085 \\ 

\emph{original} & 0.6553 $\pm$ 0.0425 & 0.4565 $\pm$ 0.0659 & 0.4381 $\pm$ 0.0701 \\ 

\emph{subq} & 0.6154 $\pm$ 0.0619 & 0.4130 $\pm$ 0.0645 & 0.3656 $\pm$ 0.0956 \\ 

\emph{pleural} & 0.7119 $\pm$ 0.0410 & 0.3783 $\pm$ 0.0525 & 0.3665 $\pm$ 0.0483 \\ 

\emph{merlin} & 0.7076 $\pm$ 0.0436 & 0.4261 $\pm$ 0.0295 & 0.4183 $\pm$ 0.0299 \\ 

\emph{subq+pleural} & 0.6303 $\pm$ 0.0560 & 0.3652 $\pm$ 0.0374 & 0.3204 $\pm$ 0.0381 \\ 

\emph{pleural+merlin} & \textbf{0.7742 $\pm$ 0.0648} & 0.5261 $\pm$ 0.1178 & 0.5040 $\pm$ 0.1467 \\ 

\emph{straightened pleural+merlin} & 0.7642 $\pm$ 0.0401 & \textbf{0.5348 $\pm$ 0.0928} & \textbf{0.5166 $\pm$ 0.1016} \\ 

\hline

\end{tabular}
}
\end{table}

% \vspace{-1.5em}

\section{Results and Discussions}		

Table~\ref{tab:video_based_classification} shows the mean and standard deviation of the video-based severity scoring metrics, obtained by evaluating on the held-out test set using the models from the five independent runs. The two models with \emph{pleural+merlin} input achieve the highest scores on all metrics, with the straightened version performing the best overall. The accuracy with the \emph{pleura} is lower than the \emph{subq} input, but combining the two gives the worst accuracy. The latter counter-intuitive result may be because the \emph{subq} and \emph{pleura} represent distinct diagnostic characteristics that AI may struggle to jointly model without seeing correlations in Merlin's space. Performance on the \emph{original} image is inferior to the \emph{pleural+merlin} image, perhaps because eliminating \emph{subq} complexity makes it easier for the model to focus on the lung region to make diagnosis, as seen in Fig~\ref{fig:gradcam_results}. Individually, \emph{merlin} has the best scores compared to \emph{subq} and \emph{pleura}.  Combining \emph{pleural+merlin} significantly improves the diagnostic accuracy of the model. The macro average RoC plots and AUC of the trained models are shown in Fig~\ref{fig:clr_roc_plot}.

Fig.~\ref{fig:gradcam_results} depicts the various masked images of a frame from a test video. The Grad-CAM \cite{Selvaraju2016Grad-CAM:Localization} visualization on the first video frame of the respective input trained model is also shown. We observe that both of the \emph{pleural+merlin} models focused on the Pleural line and B-line artifacts, whereas the \emph{original}-image-input model focused on the SubQ region. The combination of \emph{pleura+merlin} helped the model to focus on B-lines artifacts better than \emph{merlin} alone. For this test video, all models except \emph{subq} and \emph{subq+pleura} correctly predicted the severity scores, suggesting that the \emph{subq} wasn't informative to predict the diagnosis for this test video.

% We randomly selected four images for which CL trained model appeared to be looking at the correct locations (pleural line and A-line \& B-line artifacts), and we observe that in all these four cases CE trained model was basing its predictions on non-lung tissue. Refer to Fig.~\ref{fig:gradcam_results} for the model's layer-2 Grad-CAM \cite{Selvaraju2016Grad-CAM:Localization} visualization of these four images which correspond to the four severity scores. For this test images, the model correctly predicted the severity scores of 3 on all input types except subq and subq+pleura. Which suggests that the contrastive learning objective lead to learning better discriminative features.

\vspace{-1.5em}

\newlength{\width}
\setlength{\width}{0.75 in}
\newlength{\height}
\setlength{\height}{0.68 in}

% \begin{figure}[!h] 
\begin{figure}[!ht] 
\centering

\setlength{\tabcolsep}{1pt} %Column spacing
\def\arraystretch{0.5} %row spacing

% Next two lines force all the columns to be the exact same width:
\newcolumntype{C}{>{\centering\arraybackslash}m{\width}<{}}
\newcolumntype{F}{>{\centering\arraybackslash}m{0.15\width}<{}}
% \resizebox{\textwidth}{!}{
\resizebox{\columnwidth}{!}{
\begin{tabular}{F CC CC CC C}

&
\subfloat{\emph{original}} &
\subfloat{\emph{subq}} &
\subfloat{\centering \emph{subq+\newline pleural}} &
\subfloat{\emph{pleural}} &
\subfloat{\emph{merlin}} &
\subfloat{\emph{pleural+ \newline merlin}} &
\subfloat{\centering \emph{straightened} \newline \emph{pleural+ \newline merlin}} \\

\rotatebox[origin=c]{90}{\centering grey} &
\subfloat{\includegraphics[height = \height, width = \width]{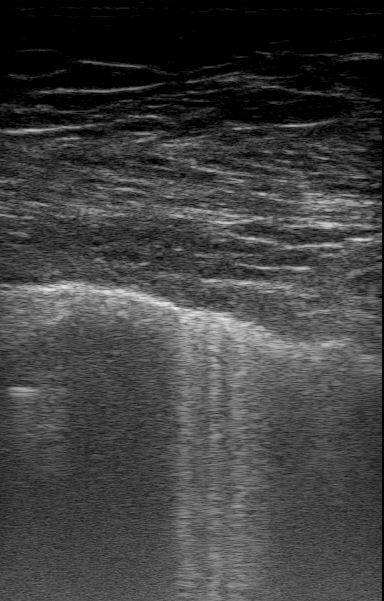}} &
\subfloat{\includegraphics[height = \height, width = \width]{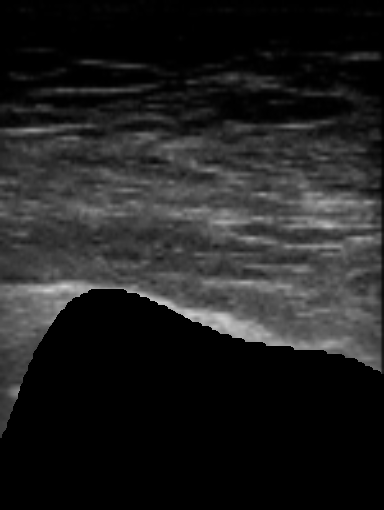}} &
\subfloat{\includegraphics[height = \height, width = \width]{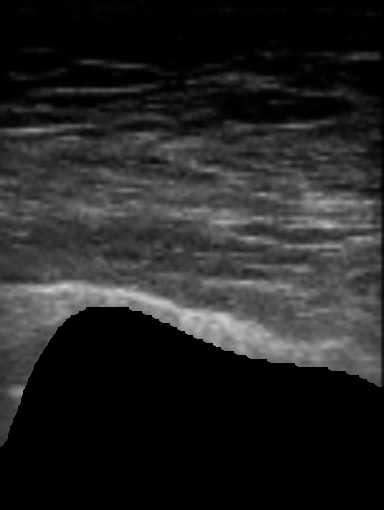}} &
\subfloat{\includegraphics[height = \height, width = \width]{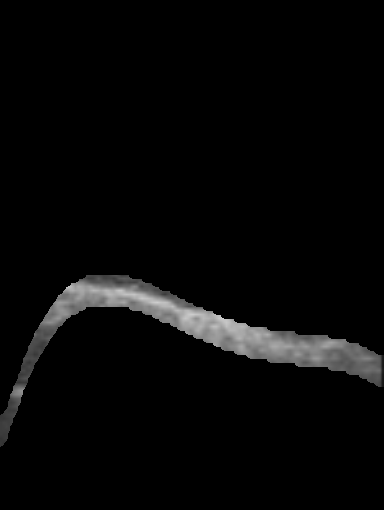}} &
\subfloat{\includegraphics[height = \height, width = \width]{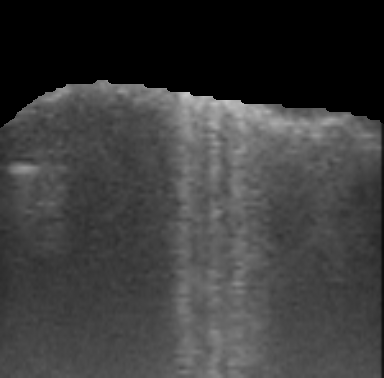}} &
\subfloat{\includegraphics[height = \height, width = \width]{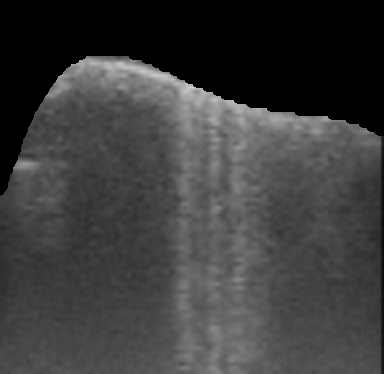}} &
\subfloat{\includegraphics[height = \height, width = \width]{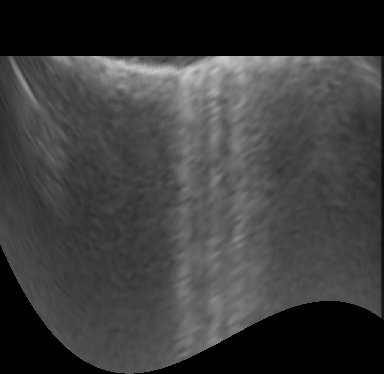}} \\

\raisebox{0.4cm}{\rotatebox[origin=c]{90}{\parbox{\height}{\centering Grad CAM}}} &
\subfloat{\includegraphics[height = \height, width = \width]{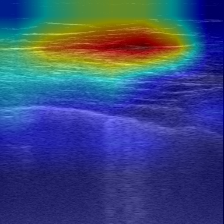}} &
\subfloat{\includegraphics[height = \height, width = \width]{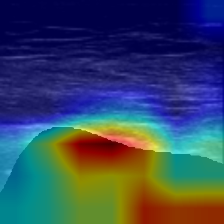}} &
\subfloat{\includegraphics[height = \height, width = \width]{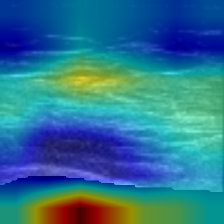}} &
\subfloat{\includegraphics[height = \height, width = \width]{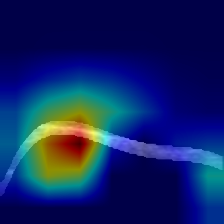}} &
\subfloat{\includegraphics[height = \height, width = \width]{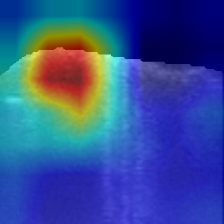}} &
\subfloat{\includegraphics[height = \height, width = \width]{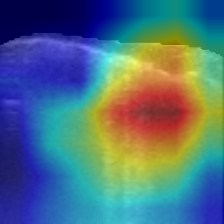}} &
\subfloat{\includegraphics[height = \height, width = \width]{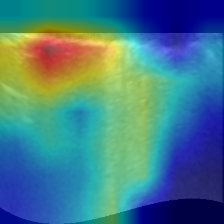}} \\ [-1ex]

% \raisebox{0.4cm}{\rotatebox[origin=c]{90}{\parbox{\height}{\centering CL RN50}}} &
% \subfloat{\includegraphics[height = \height, width = \width]{clr_R6_score-0_[ Label = 0; Pred = 0 ]_cam.png}} &
% \subfloat{\includegraphics[height = \height, width = \width]{clr_R1_score-1_[ Label = 1; Pred = 1 ]_cam.png}} &
% \subfloat{\includegraphics[height = \height, width = \width]{clr-R6_score-2_[ Label = 2; Pred = 2 ]_cam.png}} &
% \subfloat{\includegraphics[height = \height, width = \width]{clr_R1_score-1_[ Label = 1; Pred = 1 ]_cam.png}} &
% \subfloat{\includegraphics[height = \height, width = \width]{clr-R6_score-2_[ Label = 2; Pred = 2 ]_cam.png}} &
% \subfloat{\includegraphics[height = \height, width = \width]{clr_R1_score-3_[ Label = 3; Pred = 3 ]_cam.png}} \\

\end{tabular}

}
\caption{
\small Grad-CAM \cite{Selvaraju2016Grad-CAM:Localization} visualization of layer 4 of the trained model on the various masked test images (B-mode grey). We observe that the model trained on \emph{pleural+merlin} bases the predictions predominantly on the pleural line and B-line artifacts, whereas the \emph{original} image trained model predominantly bases the predictions on the subcutaneous tissues above the pleural line.}
\label{fig:gradcam_results}
% \end{figure*}
\end{figure}

\vspace{-3.5em}

\section{Conclusion}
We highlighted the potential advantages of an HFL probe over the commonly used curvilinear probe in pulmonary ultrasound. We discussed the significance of having a well-imaged pleural line in addition to pleural artifacts, such as B-lines and A-lines, suggesting that carrying out AI analysis of the pleural line using linear probe could provide new avenues for carrying out challenging diagnoses. We demonstrated the diagnostic characteristics of the subcutaneous, pleura, and Merlin's-space regions of linear-probe ultrasound. From our experiments we draw that on masking out the subcutaneous region and retaining the detailed pleura along with Merlin's space has better diagnostic prowess.

% The contrastive learning leads to better generalisation and learning of discriminative features. 
% We demonstrated that contrastive learning is robust to label noise in a weakly suprevised learning setup leading to better generalization. 

\subsubsection{Acknowledgements}
This present work was sponsored in part by US Army Medical contract W81XWH-19-C0083.
% % This work used the Extreme Science and Engineering Discovery Environment (XSEDE), which is supported by National Science Foundation grant number ACI-1548562. Specifically, it used the Bridges system, which is supported by NSF award number ACI-1445606, at the Pittsburgh Supercomputing Center (PSC).
% 
We are pursuing intellectual property protection. 
Galeotti serves on the advisory board of Activ Surgical, Inc. He and Rodriguez are involved in the startup Elio AI, Inc.

\bibliographystyle{splncs04}
\bibliography{references}

\end{document}